\documentclass{pasa}

\title[A search for predicted methanol masers]{A sensitive search for predicted methanol maser transitions with the Australia Telescope Compact Array}
\author[Chipman et al.]{A. Chipman$^1$, S.P. Ellingsen$^1$, A.M.Sobolev$^2$ and D.M. Cragg$^3$\\
\affil{$^1$School of Physical Sciences, Private Bag 37, University of Tasmania, Hobart 7001, Australia}
\affil{$^2$Ural Federal University, Lenin Avenue 51, 620000 Ekaterinburg, Russia}
\affil{$^3$School of Chemistry, 19 Rainforest Walk, Clayton Campus, Monash University, Victoria 3800, Australia}}

\jid{PASA}
\doi{10.1017/pas.\the\year.xxx}
\jyear{\the\year}

\usepackage[authoryear]{natbib}
\bibpunct{(}{)}{;}{a}{}{,}
\usepackage{aas_macros}
\usepackage{hyperref} 
\hypersetup{colorlinks,citecolor=blue,linkcolor=blue,urlcolor=blue}

\newcommand{\kms}{$\mbox{km~s}^{-1}$}
\newcommand{\ionhy}{H{\sc ii}}
\newcommand{\arcsec}{\hbox{$^{\prime\prime}$}}

\newcommand{\lta}{\raisebox{-0.6ex}{$\,\stackrel
{\raisebox{-.2ex}{$\textstyle <$}}{\sim}\,$}}

\begin{document}

\begin{abstract}
We have used the Australia Telescope Compact Array (ATCA) to search for a number of centimetre wavelength methanol transitions which are predicted to show weak maser emission towards star formation regions.  Sensitive, high spatial and spectral resolution observations towards four high-mass star formation regions which show emission in a large number of class II methanol maser transitions did not result in any detections.  From these observations we are able to place an upper limit of \lta1300K on the brightness temperature of any emission from the $3_{1}$A$^{+}$--$3_{1}$A$^{-}$, $17_{-2}$--$18_{-3}$~E ($v_t=1$), $12_{4}$--$13_{3}$~A$^{-}$, $12_{4}$--$13_{3}$~A$^{+}$  and $4_{1}$A$^{+}$--$4_{1}$A$^{-}$ transitions of methanol in these sources on angular scales of 2\arcsec.  This upper limit is consistent with current models for class~II methanol masers in high-mass star formation regions and better constraints than those provided here will likely require observations with next-generation radio telescopes.
\end{abstract}


\maketitle

\section{INTRODUCTION }
\label{sec:intro}
Methanol is an asymmetric top molecule with hindered internal rotation \citep{Lees73}, a combination which produces a rich rotational spectrum at centimetre and millimetre wavelengths.  Methanol is also a relatively common molecule in some interstellar environments. Two different types of methanol masers are observed in astrophysical environments, known as class I and class II.  Class I methanol masers are produced by outflows and other low-velocity shocks, the most common transitions are the 36 and 44 GHz and the emission is distributed over regions of up to 1 pc \citep[e.g.][]{Voronkov+14}. Class II methanol masers are produced by far infrared and submillimetre radiation, they are found close to young high-mass stars and the most common transitions are at 6.7 and 12.2 GHz.  The empirical division of the maser transitions is consistent with maser modelling which shows that the class~I masers arise in regions where collisional processes dominate \citep[e.g.][]{Sobolev+07,McEwen+14}, while the class~II masers are radiatively pumped \citep[e.g.][]{Sobolev+94,Cragg+05}.  Class I methanol masers are predominantly associated with high-mass star formation regions, but have been observed towards low-mass star formation regions \citep{Kalenskii+10}, supernova remnants \citep{Pihlstrom+14} and starburst galaxies \citep{Ellingsen+14,Chen+15}.  In contrast, class II methanol masers are only observed associated with high-mass star formation regions \citep{Breen+13b}.

To date more than 30 different class II methanol maser transitions have been observed \citep[see][and references therein]{Ellingsen+12}.  However, compared to the common 6.7 and 12.2~GHz class II methanol transitions, the others are much less common, and where they are observed are weaker than the 6.7 and 12.2 GHz emission \citep[e.g.][]{Caswell+00,Ellingsen+04,Cragg+04,Ellingsen+11a}.  The output from the methanol maser models includes predictions of the brightness temperature for each of the different transitions.  The strong, common class II methanol maser transitions (such as the 6.7 and 12.2~GHz) are predicted to produce intense maser emission for a wide range of gas temperatures, densities and methanol column densities \citep{Sobolev+97a,Cragg+05}.  In contrast, the rarer, weaker methanol transitions are typically inverted over a much narrower range of physical conditions, and the majority of transitions are not expected to be masers.  So the presence (or absence) of the rarer, weaker transitions then provides more stringent tests of maser models \citep{Cragg+04}.  The relationship between the measured flux density $S$ for emission and the brightness temperature $T_b$ of a source is given by the equation
\begin{equation} \label{eqn:bright}
  S[{\rm Jy}]  =  5.65 \times 10^{-7}\  \nu^2[{\rm GHz}]\  \theta^2[{\rm arcsec}]\  T_b[{\rm K}]
\end{equation}
where $\nu$ is the frequency of the observations and $\theta$ is the angular scale of the emission region.  For a source to be detectable at a particular frequency requires either a high brightness temperature, or emission over a large angular scale.  Maser sources generally fall into the first of these categories with strong 6.7 and 12.2 GHz methanol masers having brightness temperatures in excess of 10$^{10}$K \citep[e.g.][]{Menten+92}.  The majority of maser searches are made with single-dish telescopes with moderate (around 1 Jy) sensitivity limits, however, while these are efficient for detecting strong masers, they do not place strong upper limits on moderate or low brightness temperature masers (\lta 10$^8$ K).  Sensitive, high resolution observations of some rare class~II methanol maser transitions have shown that masers with brightness temperatures of less than 10$^8$K are likely present in some sources \citep{Krishnan+13}.  So rigorous tests of maser models require, sensitive, high resolution observations in order to detect, or limit moderate and low brightness temperature masers. The Australia Telescope Compact Array (ATCA), is an ideal instrument for undertaking such observations and in this paper we report the results of a search for a number of methanol maser transitions which are predicted to exhibit weak maser emission under some physical conditions.

In order to test and refine maser models and potentially the physical conditions in the observed star formation regions, we selected a number of potential class II methanol maser transitions in the frequency range 5--9~GHz.   A search for potential new maser transitions is a compromise between the number of different sources observed and the sensitivity of the observations.  For the primary target transitions we made observations of four high-mass star formation regions which are known to exhibit moderately strong emission from a range of different class II methanol maser transitions.  For the secondary target transitions we made observations of a single source G\,345.01$+$1.79, as there are more class II methanol maser transitions detected toward that source than any other \citep{Ellingsen+12}.

\section{OBSERVATIONS AND DATA REDUCTION}

We used the ATCA in the 6A configuration (baselines ranging from 337 to 5939m) to undertake sensitive observations of 5 different methanol (CH$_{3}$OH) transitions in the 5--9 GHz frequency range.  The observations were made over a period of 13 hours on 2000 August 30. The primary target transitions were the $12_{4}$--$13_{3}$~A$^{-}$ and $12_{4}$--$13_{3}$~A$^{+}$ transitions, which have rest frequencies of 7.682232(50) and 7.830864(50) GHz  respectively \citep[][the numbers in brackets represent the uncertainty in the least significant figures]{Tsunekawa+95}.  These two transitions were observed towards four high-mass star formation regions (G\,339.88-1.26, G\,345.01+1.79, NGC6334F and W48), which all exhibit strong class~II methanol maser emission in multiple transitions.  The pointing centre for the observations of each source and the peak velocity of the class~II methanol maser transitions is given in Table~\ref{tab:obs}. For each transition, observations consisted of a series of scans of 5 minutes for each target source spread over a range of hour angles.   Each scan on a target source was both preceded and followed by a 1 minute scan on a nearby phase calibrator.  The total observing time for each of the four target maser sources is around 45 minutes for each of the primary transitions.

Observations were also undertaken of secondary transitions $3_{1}$A$^{+}$--$3_{1}$A$^{-}$, $17_{-2}$--$18_{-3}$~E ($v_t=1$) and $4_{1}$A$^{+}$--$4_{1}$A$^{-}$ which have rest frequencies of 5.0053207(6), 6.181589(153) and 8.341622(4)~GHz respectively \citep{Breckenridge+95,Xu+97}.  These three transitions were observed only towards the G\,345.01+1.79 star formation region with onsource durations of 4.2 minutes for each transition.  Of the 5 transitions observed, the only one which has been detected astronomically is the $3_{1}$~A$^{+}$--$3_{1}$~A$^{-}$, for which \citet{Robinson+74} report weak, broad emission towards Sgr\,B2.

\begin{table*}
\caption{Class II methanol maser sources observed in the $12_{4}$--$13_{3}$~A$^{-}$ and $12_{4}$--$13_{3}$~A$^{+}$ transitions, with the measured RMS noise for each transition respectively.  Distance references $^{\dagger}$ \citet{Krishnan+15} ; $^{\mathsection}$ \citet{Green+McClure11}$^{\ddagger}$ \citet{Wu+14} }
\begin{center}
\begin{tabular}{@{}lllrrccr@{}}
\hline\hline
                   & {\bf Right}          &                            &  {\bf Peak}   & {\bf Velocity} & $12_{4}$--$13_{3}$ A$^{-}$ &  $12_{4}$--$13_{3}$ A$^{+}$ &  \\
                   & {\bf Ascension}  & {\bf Declination} & {\bf Velocity} & {\bf Range}  & \multicolumn{2}{c}{\bf RMS for transition}                              & {\bf Distance} \\
{\bf Source}& {\bf (J2000)}      & {\bf (J2000)}        & {\bf (\kms)}  & {\bf \kms}      & \multicolumn{2}{c}{\bf (mJy beam$^{-1}$)}                            & {\bf (kpc)}        \\
\hline%
G\,339.88$-$1.26 & 16:52:04.68  & $-$46:08:34.4 & $-$39.0 & $-$44 -- $-$34 & 33 & 27 & 2.1$^{\dagger}$ \\
G\,345.01$+$1.79 & 16:56:47.58  & $-$40:14:25.9 & $-$21.0 & $-$24 -- $-$14 & 35 & 29 & 2.0$^{\mathsection}$ \\
NGC6334F            & 17:20:53.45  & $-$35:47:00.5 & $-$10.0 & $-$14 -- $-$4 & 36 & 29 & 1.35$^{\ddagger}$ \\
W48                       & 19:01:46.00  & $+$01:12:29.0 & $+$45.0 & $+$38 -- $+$48 & 35 & 39 & 3.27$^{\ddagger}$ \\
\hline\hline
\end{tabular}
\end{center}
\label{tab:obs}
\end{table*}

The correlator was configured to the FULL\_4\_1024\_POL mode which has 1024 spectral channels across a 4 MHz bandwidth, with all polarization products recorded.  This corresponds to a velocity coverage of approximately 155~\kms\/ at a frequency of 7.7~GHz and velocity resolution of approximately 0.18~\kms.  The data were reduced using {\sc miriad}, applying the standard data reduction procedures for ATCA observations.  The primary flux density calibration was with respect to the standard calibrator PKS\,B1934-638 using the in-built scale within {\sc miriad} \citep{Reynolds94}.  The absolute flux density calibration is expected to be accurate to better than 10\%. PKS\,B1934-638 was also used for bandpass calibration.

\section{RESULTS}

\begin{figure*}
\begin{center}
\includegraphics[width=17cm]{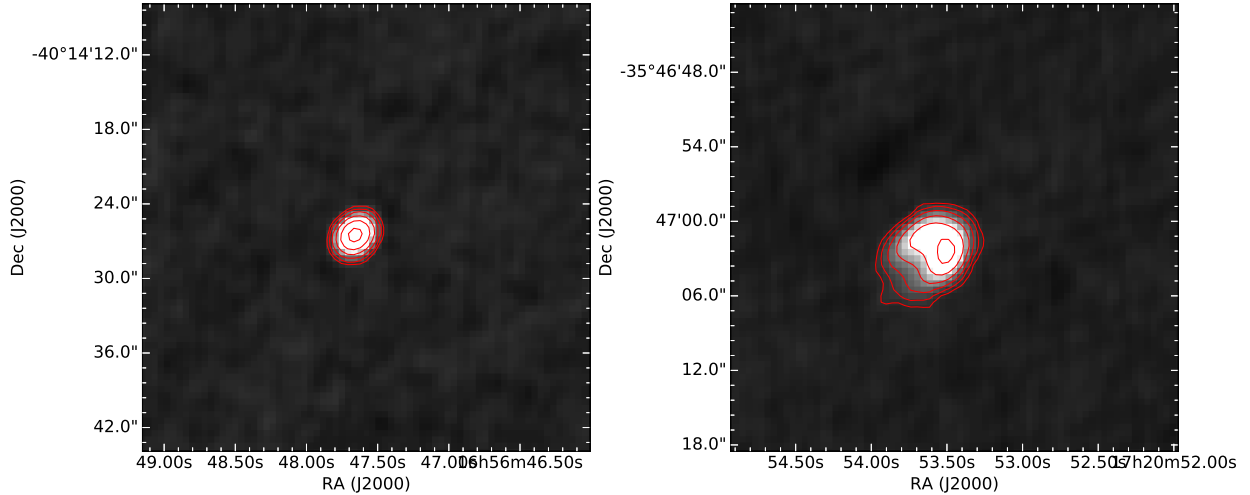}
\caption{Radio continuum emission at 7.8~GHz for G\,345.01+1.79 (left) and NGC6334F (right).  For both sources contour levels are at 2.5, 5, 10, 20, 40 and 80\% of the image peak which is 217 and 776 mJy beam$^{-1}$ for G\,345.01+1.79 and NGC6334F respectively.  The RMS noise level as are 1.4 and 4.1 mJy beam$^{-1}$ for G\,345.01+1.79 and NGC6334F respectively.  The synthesised beam for these observations was approximately 2.0\arcsec $\times$ 1.8\arcsec.}\label{fig:images}
\end{center}
\end{figure*}

We imaged each of the four sources using the calibrated 4 MHz datasets for the $12_{4}$--$13_{3}$~A$^{-}$ and $12_{4}$--$13_{3}$~A$^{+}$ transitions.  Three of the four sources (all except G\,339.88-1.26) have detectable radio continuum emission in both transitions.  For these sources we undertook self-calibration using a model formed from the cleaned images of the radio continuum emission.  We found that a single iteration of phase-only self-calibration significantly improved the signal-to-noise ratio in the continuum images and we then used continuum subtraction to extract a spectral line dataset.  We imaged a 10~\kms\/ velocity range covering all the class II methanol maser emission in each source at a velocity resolution of 0.2~\kms (see Table~\ref{tab:obs}).  Inspection of the image cubes for each of the sources did not reveal any emission in either of the two primary target transitions in any of the four source observed.  The RMS noise level in 0.2~\kms\/ spectral channels were in the range 27 -- 37~mJy beam$^{-1}$ for all sources and are listed in Table~\ref{tab:obs}.  Figure~\ref{fig:images} shows the continuum images formed from the $12_{4}$--$13_{3}$~A$^{+}$ dataset after self-calibration for G\,345.01+1.79 and NGC6334F.  We do not show the continuum image for W48 because as it is very close to the equator, the ATCA as an East-West synthesis array has a very elongated synthesised beam and does not produce good images.  Comparison of the images in Figure~\ref{fig:images} with previous more sensitive observations of these sources at centimetre wavelengths \citep[e.g.][]{Ellingsen+05} shows similar structure and intensity.  This demonstrates that despite both of these transitions lying outside the nominal frequency range of the ATCA receiver system (at the time of the observations) the system performed well and the non-detection of the spectral line emission is robust.

The three secondary transitions ($3_{1}$A$^{+}$--$3_{1}$A$^{-}$, $17_{-2}$--$18_{-3}$~E ($v_t=1$) and $4_{1}$A$^{+}$--$4_{1}$A$^{-}$) were each observed for 4.2 minutes towards G\,345.01+1.79.  We have calibrated these data using the same approach as for the primary transitions, however, with a one-dimensional array configuration (such as the ATCA in the 6A array), it is not possible to image data from a single scan.  We produced a vector averaged spectrum at the phase centre for each of the transitions and from this we obtained upper limits (3$\sigma$) on the intensity of any emission of 0.17~Jy, 0.15~Jy \& 0.25~Jy for the $3_{1}$A$^{+}$--$3_{1}$A$^{-}$, $17_{-2}$--$18_{-3}$~E ($v_t=1$) and $4_{1}$A$^{+}$--$4_{1}$A$^{-}$ transitions respectively.  The effective field of view for the vector averaged spectrum is the size of the synthesised beam (a few arcseconds), centred on the coordinates given for G\,345.01+1.79 given in Table~\ref{tab:obs}. This is sufficient to encompass the class~II methanol masers in this source which are observed to be confined to a single compact cluster with an angular extent of less than 0.5 arcseconds \citep{Krishnan+13}.

\section{DISCUSSION}

\citet{Sobolev+97b} used the methanol maser pumping model of \citet{Sobolev+97a} to predict which transitions are likely to show maser emission under a range of different physical conditions.  In total 6 different pumping model input parameters were varied and the results for 10 different combinations are listed in table~1 of \citet{Sobolev+97b}.  The following model parameters were varied, the gas temperature ($T_{\rm kin}$), the gas density ($n_{\rm H}$), the maser beaming factor ($\epsilon$), the dilution of continuum emission from a background \ionhy\/ region ($W_{\rm H_{II}}$), the dust temperature ($T_{\rm d}$) and the methanol specific column density $N_{\rm M}/\Delta V$.  The $12_{4}$--$13_{3}$~A$^{-}$ and $12_{4}$--$13_{3}$~A$^{+}$ transitions were predicted to exhibit weak maser emission (brightness temperatures of 10$^4$ -- 10$^5$K) in 6 of the 10 models, more than any of the other methanol transitions at frequencies $<$ 10~GHz which had not previously been detected.   The most recent published class~II methanol maser pumping model is that of \citet{Cragg+05}, which is based on the earlier work of \citeauthor{Sobolev+97a}, but incorporates a number of improvements and refinements, including new collisional data and more energy levels.  The \citet{Cragg+05} model also suggests that the $12_{4}$--$13_{3}$~A$^{-}$ and $12_{4}$--$13_{3}$~A$^{+}$ transitions are good candidates for weak masers, with 6 of the 8 sets of input physical parameters they trialled yielding masers with brightness temperatures $>$ 10$^3$K, with three of them having brightness temperatures in excess of 10$^4$K.

The four sources targeted for the current search are a subset of the 25 high-mass star formation regions which have been observed to show maser emission from the 
107~GHz $3_{1}$--$4_{0}$~A$^{+}$ transition.  Table~5 of \citet{Ellingsen+11a} summarises the observed intensity, or measured upper limit for a dozen different class~II methanol maser transitions towards these sources and shows that G\,339.88$-$1.26, G\,345.01$+$1.79, NGC6334F and W48 have been detected in 6, 10, 8 and 5 of these dozen transitions respectively.  Comparing the class~II methanol maser transitions observed in our four target sources with the results in table~2 of \citet{Cragg+05}, they best match model 3, as it is the only model that predicts moderately strong 37.7 GHz methanol masers, which all these sources have.  This model has $T_{\rm kin}$ = 30K, $T_{\rm d}$=175K, $n_{\rm H}$ = 10$^7$ cm$^{-3}$, $N_{\rm M}/\Delta V$ = $3 \times 10^{11}$ cm$^{-3}$s and $W_{\rm H_{II}}$ = 0.002.  This model predicts brightness temperatures of $\sim 6.3 \times 10^{3}$~K for the $12_{4}$--$13_{3}$~A$^{-}$ and $12_{4}$--$13_{3}$~A$^{+}$ transitions.

Table~\ref{tab:obs} shows that our most sensitive observations were of the $12_{4}$--$13_{3}$~A$^{+}$ transition at 7.83~GHz for which we obtained RMS noise levels in 0.2~\kms\/ spectral channels of $<$ 30 mJy beam$^{-1}$ for three of the four sources (we had less time onsource for W48, so the RMS noise is higher).  We can be confident that there is no emission stronger than 0.150 Jy beam$^{-1}$ (5-$\sigma$) in this transition in the observed sources and from equation~\ref{eqn:bright} that places an upper limit of $\sim 1100$K on the brightness temperature of any emission on angular scales of 2\arcsec\/ (the angular resolution of our observations). This is significantly less than the predicted brightness temperature in the \citet{Cragg+05} model, however, maser emission generally occurs on much smaller angular scales than the 2\arcsec\/ resolution of our synthesised beam.  Very long baseline interferometry observations of class~II methanol masers \citep[e.g.][]{Menten+92,Minier+02a,Harvey-Smith+06} show that the 6.7~GHz class II methanol masers generally have angular sizes in the range 5--50 milliarcseconds.  Rearranging equation~\ref{eqn:bright} for the $12_{4}$--$13_{3}$~A$^{+}$ transition, our 0.15 Jy beam$^{-1}$ upper limit on the flux of the emission from this transition implies $\theta^2 T_b < 4330$ K arcsecond$^2$.  This corresponds to upper limits on the brightness temperature of $1.7 \times 10^6$K for emission on scales of 50 milliarcseconds or $1.7 \times 10^8$K for emission on scales of 5 milliarcseconds.  These limits are both more than a order of magnitude higher than the predictions of any of the models of either \citet{Sobolev+97b} or \citet{Cragg+05} for this transition.  

The observations of the $12_{4}$--$13_{3}$~A$^{-}$ transition at 7.62~GHz are approximately 20\% less sensitive than those for the $12_{4}$--$13_{3}$~A$^{+}$ transition.  The lower sensitivity is believed to be due to poorer receiver performance at the frequency of that transition (at the time of our observations).  The affect of this is that the limits for the $12_{4}$--$13_{3}$~A$^{-}$ transition are 20\% higher, however, this makes no significant difference to the conclusions we can draw.  

It should be noted that the analysis undertaken here utilises data with a 0.2~\kms\/ velocity resolution.  This is close to optimal for Galactic maser work, as typical line widths are around 0.5~\kms.  Averaging several spectral channels would yield lower RMS noise levels in spectral line cubes, however, the resulting limits would be misleading, as they would not account for the dilution of narrow line emission.

From equation~\ref{eqn:bright} we can estimate the minimum sensitivity for observations of the transitions targeted in this search that would be required to provide a stringent test of methanol maser pumping models.  If we take a maximum likely size for the angular scale of the maser emission as 50 milliarcseconds then a maser brightness temperature of 10$^4$K will result in a flux density of around 1~mJy.  Robust detection of emission at this level in a 0.2~\kms\/ spectral channel would require observations about 2 orders of magnitude more sensitive than those we report here. This level of sensitivity could be achieved with very long (of order days) integrations on sensitive telescopes with large collecting areas (e.g. the GBT, JVLA or FAST), however, it more realistically lies within the realm of a future SKA survey of line emission in Galactic high-mass star formation regions.
 
\section{CONCLUSION}

We have searched for emission from a number of centimetre wavelength methanol transitions which are predicted to show weak maser emission towards high-mass star formation regions under some physical conditions.  Our observations did not detect emission from any of the transitions in any of the four sources observed with upper limits of approximately 0.15~Jy.  The observations reported here are approximately an order of magnitude more sensitive and an order of magnitude higher angular resolution than typical single dish searches for maser transitions and so they do represent a significant improvement on many searches for weak, rare class~II methanol masers \citep[e.g.][]{Ellingsen+03,Cragg+04}.  Previous observations have demonstrated that more sensitive observations can lead to the detection of weak maser lines, for example \citet{Sobolev+07} detected weak 23.1~GHz class II methanol masers after earlier unsuccessful searches by \citet{Cragg+04}.  To significantly improve on the limits placed on the $3_{1}$A$^{+}$--$3_{1}$A$^{-}$, $17_{-2}$--$18_{-3}$~E ($v_t=1$), $12_{4}$--$13_{3}$~A$^{-}$, $12_{4}$--$13_{3}$~A$^{+}$  and $4_{1}$A$^{+}$--$4_{1}$A$^{-}$ transitions of methanol by the current observations will require next-generation radio astronomy facilities.

\begin{acknowledgements}
The Australia Telescope Compact Array is part of the Australia Telescope which is funded by the Commonwealth of Australia for operation as a National Facility managed by CSIRO. This research has made use of NASA's Astrophysics Data System Abstract Service.  AC was funded by a University of Tasmania Dean's Summer Research scholarship.  AMS has been supported by the Ministry of Education and Science of the Russian Federation within the framework of state work ``Organization of Execution of Scientific Research''. This work was also supported by Government of the Russian Federation Act 211, contract no 02.A03.21.0006.
\end{acknowledgements}

\bibliographystyle{pasa-mnras}
\bibliography{references}

\end{document}